# Economic Accelerator with Memory: Discrete Time Approach

**Valentina V. Tarasova**

Lomonosov Moscow State Business School, Lomonosov Moscow State University,

Moscow 119991, Russia; E-mail: v.v.tarasova@mail.ru

**Vasily E. Tarasov**

Skobeltsyn Institute of Nuclear Physics, Lomonosov Moscow State University,

Moscow 119991, Russia; E-mail: v.v.tarasov@bk.ru; tarasov@theory.sinp.msu.ru

**Abstract.** Accelerators with power-law memory are proposed in the framework of the discrete time approach. To describe discrete accelerators we use the capital stock adjustment principle, which has been suggested by Matthews. The suggested discrete accelerators with memory describe the economic processes with the power-law memory and the periodic sharp splashes (kicks). In continuous time approach the memory is described by fractional-order differential equations. In discrete time approach the accelerators with memory are described by discrete maps with memory, which are derived from the fractional-order differential equation without approximations. In order to derive these maps we use the equivalence of fractional-order differential equations and the Volterra integral equations.

**Keywords:** accelerator, power-law memory, macroeconomics, Matthews' capital stock adjustment principle, discrete map with memory

**JEL Classification:** C00; C02; C65; E00; E20

1. Introduction

One of the basic concepts of macroeconomics is the accelerator [1, 2, 3, 4]. Accelerator equations can be represented in the frameworks of the discrete time and continuous time approaches. We consider an exact correspondence between these two approaches for economic processes with power-law memory. Initially we prove that the discrete accelerator equations, which contain standard finite differences, can be derived from the differential equation with periodic kicks. Using the generalization of these equations, which take into account power-law memory, we derive the discrete accelerator equations with memory in the form of discrete maps with memory.

## 2. Accelerator without memory

In continuous terms, the simplest equation of the accelerator [1, p. 62] is the continuous linear form without memory

$$\frac{dY(t)}{dt} = \frac{1}{v} \cdot I(t), \tag{1}$$

where $dY(t)/dt$ is the rate of output (income), $I(t)$ is the rate of induced investment, and v is a positive constant, which is called the investment coefficient [1, p. 62] that indicates the power of the accelerator. The coefficient v is also called the accelerator coefficient or the capital intensity of the income growth rate [2, p. 91], where $1/v$ is the capital productivity incremental or marginal productivity of capital [2, p. 91]. Equation (1) means that the induced investment is here a constant proportion of the current rate of change of output.

In discrete time approach, the accelerator without memory is to be written [1, p. 63] in the linear form by the equation

$$Y_n - Y_{n-1} = \frac{T}{v} \cdot I_n. \tag{2}$$

Here $Y_n = Y(nT)$ and $I_n = I(nT)$, where T is a positive constant indicating the time scale. If T=1, then t=n and $Y_n = Y_t$. In this case, equation (2) has the form $I_t = v \cdot (Y_t - Y_{t-1})$. Equation (2) means that induced investment depends on the current change in output [1, p. 63].

## 3. Capital stock adjustment principle

There is an alternative approach to the accelerator equation, which is proposed by Matthews in the form of the capital stock adjustment principle [3].

Let us consider the capital stock $K(t)$, which can be taken as varying, and so the level of net investment $I(t)$ depends on $K(t)$ [4, p. 68]. The investment $I(t)$ must depend on profits, both as an indicator of profitability of production (or the level of demand), and as one of the sources of funds available to finance investment. In general case, we can take income $Y(t)$, instead of profits, as an indicator of the level of demand and of the availability of finance. On this approach, we can write the investment function $I=I(Y,K)$, where we ignored the influence of the interest rate. In the linear case, the investment function is linear in $Y(t)$ and $K(t)$, and we have

$$I(Y(t), K(t),) = a \cdot Y(t) - b \cdot K(t), \tag{3}$$

where a and b are positive coefficient of the investment function. This case corresponds to the capital stock adjustment principle proposed by Matthews [3].

We can approach to the accelerator equation by starting from the linear investment function (3) on the basis of the Matthews' capital stock adjustment principle [3, 4]. Considering the particular case of this principle in which b = 1 and T=1, we have [4, p. 73] the equation

$$a \cdot Y_t = K_t + I_t = K_{t+1}. \tag{4}$$

Then we have $K_{t+1} = a \cdot I_t$ and $K_t = a \cdot I_{t-1}$ that give

$$I_t = K_{t+1} - K_t = a \cdot (Y_t - Y_{t-1}). \tag{5}$$

As a result, the acceleration equation is obtained as a particular case (a=v, b=1, T=1) of the capital stock adjustment principle.

Let us consider the Harrod-Domar model, which is translatable into period terms. The variables for a sequence of periods t=0, T, 2T, ... are output (income) $Y_n$ as the flow in the period t=nT, the capital stock $K_n$ timed at the beginning of the period, and the investment in period

t=nT is given by $I_n$ [4, p. 204]. In the fixed-coefficients version of the Harrod-Domar model, the "investment = saving" equation (equation 1 of Section 11.4 in [4, p. 204]) has the form

$$K_{n+1} - K_n = s \cdot T \cdot Y_n, \tag{6}$$

where the parameter s is the constant propensity to save. For T=1, equation (6) has the form $K_{t+1} - K_t = s \cdot Y_t$. In the continuous time approach, equation (6) is considered (equation 1 of Section 11.2 in [4, p. 199]) in the form

$$\frac{dK(t)}{dt} = s \cdot Y(t), \tag{7}$$

where dK(t)/dt is the derivative of first order of the capital stock function K(t).

This formulation can be relaxed [4, p. 204-205] by dropping explicit reference to the capital stock and the production function. Then full capacity equation (equation 4 of Section 11.4 in [4, p. 205]) has the form

$$Y_{n+1} - Y_n = \frac{T}{v} \cdot I_n, \tag{8}$$

where v is a fixed coefficient of the production function. For T=1, equation (8) has the form $I_t = v \cdot (Y_{t+1} - Y_t)$. In the continuous time approach, equation (8) is considered (equation 1 of Section 11.2 in [4, p. 199]) in the form (1), where dY(t)/dt is the derivative of first order of the income function Y(t).

The multiplier-accelerator version is a variant of (8) in which the full-capacity condition is reversed [4, p. 205] in its lag/lead interpretation to give the investment function by equation (3).

## 4. Connection between discrete and continuous time approaches

Equations (2), (8) and (6) cannot be considered as exact discrete analogs of equation (1) and (7). This is caused by that the standard finite differences, such as the forward difference $\Delta^1_{forward} Y(t) := Y(t+1) - Y(t)$, and the backward difference $\Delta^1_{backward} Y(t) := Y(t) - Y(t-1)$, do not have the same basic characteristic properties as the derivatives of first order [5, 6]. For example, the standard Leibniz rule (the product rule) is violated for these finite differences [5, 6].

Using the approach, which is suggested in [7, 8] and [9, p. 409-453], we can propose the differential equation of the accelerator that gives discrete time analogs of equations (2), (8) and (6), which corresponds to the capital stock adjustment principle. Discrete equations (2), (8) and (6) can be derived from the suggested differential equation without the use of any approximations.

Let us consider the differential equations

$$\frac{dK(t)}{dt} = s \cdot Y(t) \cdot \sum_{k=1}^{\infty} \delta\left(\frac{t}{T} - k\right), \tag{9}$$

$$\frac{dY(t)}{dt} = \frac{1}{v} \cdot I(t) \cdot \sum_{k=1}^{\infty} \delta\left(\frac{t}{T} - k\right), \tag{10}$$

where δ(z) is the Dirac delta-function, which is a generalized function [10, 11]. The delta-function has an important role in modern economics and finance [12, 13]. The delta functions describe the periodic sharp splashes (kicks). It should be noted that the generalized functions are treated as continuous functionals on a space of test functions. These functionals are continuous in a suitable topology on the space of test functions. Therefore equations (9) and (10) should be understood in a generalized sense i.e. on the space of test functions.

To derive a discrete equation from equations (9) and (10), we can use the fundamental theorem of calculus and the Newton-Leibniz formula in the form

$$\int_0^t f^{(1)}(\tau)d\tau = f(t) - f(0), \qquad (11)$$
where $f^{(1)}(\tau) := df(\tau)/d\tau$ is the derivative of first order.

Using integration of equation (9) from 0 to t, where nT<t<(n+1)T, we get the discrete equation
$$K_{n+1} = K_0 + s \cdot T \cdot \sum_{k=1}^{n} Y_k, \qquad (12)$$
where $K(0) = K_0$, and
$$Y_k := Y(k \cdot T - 0) = \lim_{\varepsilon \to 0+} Y(k \cdot T - \varepsilon), \qquad (13)$$
$$K_{n+1} := K((n+1) \cdot T - 0) = \lim_{\varepsilon \to 0+} K((n+1) \cdot T - \varepsilon). \qquad (14)$$
In equation (12), we replace $n + 1$ by n and get
$$K_n = K_0 + s \cdot T \cdot \sum_{k=1}^{n-1} Y_k. \qquad (15)$$
Subtracting equation (15) from equation (12), we obtain $K_{n+1} - K_n = s \cdot T \cdot Y_n$, which coincides with equation (6).

As a result, we can conclude that discrete time equations (6) and (8) correspond to the differential equations (9) and (10) respectively. In the continuous time approach the discrete economic accelerators describe economic processes with the periodic sharp splashes (kicks), which are represented by delta functions. We can state that the discrete accelerator (6) actually describes the economic dynamics with the periodic sharp splashes of the output or the periodic sharp splashes of propensity to save s. The discrete accelerator (8) actually describes the economic dynamics with the periodic sharp splashes of the net investment or periodic sharp splashes of capital productivity 1/v.

## 5. Continuous time accelerator with power-law memory

To take into account the power-law memory effect in acceleration principle and the Matthews capital stock adjustment principle, we can use concept of the marginal values of non-integer order [17, 18] and the accelerators of non-integer order [19]. Equations (1) and (7) can be generalized by using accelerators with memory [19], which describes the relationship between the net investment (the output) and the margin output (the capital stock) of non-integer order. In order to have the correct dimensions of economic quantities we will use the dimensionless time variable t. These generalizations of the standard accelerator equations (1) and (7), which takes into account the memory effects of the order α, can be given [19] in the form
$$(D_{0+}^\alpha Y)(t) = \frac{1}{v} \cdot I(t), \qquad (16)$$
$$(D_{0+}^\alpha K)(t) = s \cdot Y(t), \qquad (17)$$
where $D_{0+}^\alpha$ is the left-sided Caputo derivative of the order α>0, which is defined by
$$(D_{0+}^\alpha K)(t) := \frac{1}{\Gamma(n-\alpha)} \int_0^t \frac{K^{(n)}(\tau)d\tau}{(t-\tau)^{\alpha-n+1}}, \qquad (18)$$
where $\Gamma(\alpha)$ is the gamma function, $K^{(n)}(\tau)$ is the derivative of the integer order $n := [\alpha]+1$ of the function $K(\tau)$ with respect to τ: 0<τ<t. For the existence of the expression (18), the function $K(\tau)$ must have the integer-order derivatives up to the (n-1)-order, which are absolutely continuous functions on the interval [0, t]. For integer α = n, the Caputo derivatives coincide with standard derivatives [15, p. 79], [16, p. 92-93], i.e. $(D_{0+}^n K)(t) = K^{(n)}(t)$. The derivatives of non-integer order have an economic interpretation [20, 21]. It should be noted that equations (16) and (17) with α = 1 take the form (1) and (7) respectively.

The accelerator equations (16) and (17) include the standard equations of the accelerator and the multiplier, as special cases [19]. This statement can be proved by considering these equations for α=0 and α=1. Using the property $(D_{0+}^1 Y)(t) = Y^{(1)}(t)$ of the Caputo derivative [16, p. 79], formula (16) with α=1 takes give equation (1) that describes the standard accelerator. Using $(D_{0+}^0 Y)(t) = Y(t)$, equation (16) with α=0 is written as I(t)=v·Y(t), which is the equation of standard multiplier. Therefore, the concept of the accelerator with memory generalizes the concepts of the standard multiplier and accelerator [19].

Let us consider the fractional differential equations

$$(D_{0+}^\alpha Y)(t) = \frac{1}{v} \cdot I(t) \cdot \sum_{k=1}^\infty \delta\left(\frac{t}{T} - k\right), \quad (19)$$

$$(D_{0+}^\alpha K)(t) = s \cdot Y(t) \cdot \sum_{k=1}^\infty \delta\left(\frac{t}{T} - k\right), \quad (20)$$

which should be understood in a generalized sense i.e. on the space of test functions. In the framework of continuous time approach, equations (19) and (20) can be considered as accelerator equations for economic processes with memory and crises (periodic sharp splashes).

The action of the left-sided Riemann-Liouville fractional integral of the order α on equations (19) and (20) is defined on the space test functions on the half-axis by using the adjoint operator approach [14, p. 154-157]. The left-sided Riemann-Liouville fractional integration provides operation inverse [16, p. 96–97] to the left-sided Caputo fractional differentiation, which is used in equations (19) and (20). The Lemma 2.22 of [16, p. 96–97] is a basis of the equivalence of fractional differential equations and the Volterra integral equations [16, p. 199-208]. For fractional differential equations (19) and (20) this equivalence should be considered on the space of test functions [14, p. 154-157], since these equations contain the Dirac delta functions.

## 6. Discrete time accelerator with power-law memory

Let us obtain an equation of discrete accelerator with memory, which corresponds to fractional differential equation (20). For this purpose, we use Theorem 18.19 of [9, p. 444], which is valid for any positive order α>0 and which was initially suggested in [7, 8]. This theorem is based on the equivalence of fractional differential equations and the Volterra integral equations in the generalized sense, i.e. on a space of test functions. Using Theorem 18.19 of [9, p. 444], we can state that the Cauchy problem with differential equation (20) and the initial conditions $K^{(k)}(0) = K_0^{(k)}$ (k=0, 1,... , N-1), where N-1<α<N, is equivalent to the discrete equation

$$K_{n+1}^{(m)} = \sum_{k=0}^{N-m-1} \frac{T^k}{k!} \cdot K_0^{(k+m)} \cdot (n+1)^k +$$
$$\frac{s \cdot T^{\alpha-m}}{\Gamma(\alpha-m)} \cdot \sum_{k=1}^n (n+1-k)^{\alpha-1-m} \cdot Y_k, \quad (21)$$

where $Y^{(m)}(t) = d^m Y(t)/dt^m$, $Y_k^{(m)} := \lim_{\varepsilon \to 0+} Y^{(m)}(k \cdot T - \varepsilon)$, and m=0, 1, …, N-1. Equation (21) is the discrete map with memory. Equation (21) defines the accelerator with memory in the framework of discrete time approach.

We should emphasize that discrete equation (21) is derived from the fractional differential equation (20) without the use of any approximations, i.e. it is an exact discrete analog of the fractional differential equation (20). Equations (21) define a discrete map with power-law memory of the order α> 0.

For $0<\alpha<1$ (N = 1) the discrete map (21) is described by the equation

$$K_{n+1} = K_0 + \frac{s \cdot T^\alpha}{\Gamma(\alpha)} \cdot \sum_{k=1}^{n}(n + 1 - k)^{\alpha-1} \cdot Y_k. \tag{22}$$

In equation (22), the replacement n + 1 by n gives

$$K_n = K_0 + \frac{s \cdot T^\alpha}{\Gamma(\alpha)} \cdot \sum_{k=1}^{n-1}(n - k)^{\alpha-1} \cdot Y_k. \tag{23}$$

Subtracting equation (23) from equation (22), we obtain

$$K_{n+1} - K_n = \frac{s \cdot T^\alpha}{\Gamma(\alpha)} \cdot Y_n + \frac{s \cdot T^\alpha}{\Gamma(\alpha)} \sum_{k=1}^{n-1} V_\alpha(n - k) \cdot Y_k, \tag{24}$$

where $V_\alpha(z)$ is defined by $V_\alpha(z) := (z + 1)^{\alpha-1} - (z)^{\alpha-1}$. Equation (24) is a generalization of equation (6) for the case of power-law memory with $0<\alpha<1$.

. Similarly, fractional differential equation (19) gives the discrete equation

$$Y_{n+1} - Y_n = \frac{T^\alpha}{v \cdot \Gamma(\alpha)} \cdot I_n + \frac{T^\alpha}{v \cdot \Gamma(\alpha)} \cdot \sum_{k=1}^{n-1} V_\alpha(n - k) \cdot I_k, \tag{25}$$

which is a generalization of equation (8) for the case of power-law memory with $0<\alpha<1$.

Equations (24) and (25) describe the discrete analogs of accelerators with memory of the order $0<\alpha<1$. For $\alpha=1$, we can use $V_1(z) = 0$, and equations (24) and (25) give the discrete maps, which coincide with equation (6) and (8) respectively.

## 7. Conclusion

As a result, we can conclude that fractional differential equations (19) and (20) exactly correspond to discrete time equations (24) and (25) respectively. In the continuous time approach the discrete economic accelerators with memory are described by economics dynamics with memory and the periodic sharp splashes (kicks). The discrete accelerator (24) actually describes the economic processes with memory and the periodic sharp splashes of the output (or the periodic sharp splashes of the propensity to save s). The discrete accelerator (25) actually describes the economic dynamics memory and the periodic sharp splashes of the net investment (or the capital productivity 1/v).

We emphasize that there is no exact correspondence between the standard discrete and continuous time accelerators, which are described by equations (1) and (2), (6) and (7), (1) and (8). These accelerators are connected only asymptotically. In continuous time approach, the standard discrete accelerators (6) and (8) correspond exactly to the differential equations with the periodic sharp splashes (kicks), which are represented by the delta functions. Therefore the standard discrete accelerators without memory correspond to the continuous time dynamics with the periodic sharp splashes. The suggested discrete accelerators with memory, which are defined by equations (21), (24) and (25), describe the economic processes with power-law memory and periodic sharp splashes. The suggested discrete accelerators with memory can be used to describe economic and finance processes with power-law memory [22, 23, 24, 25, 26, 27, 28, 29, 30, 31, 32] in the framework of discrete time approach.